\documentclass[a4paper,final]{appolb}
\usepackage{epsfig}
\def\runhead{J.~Raufeisen}
\def\thetitle{Probing the nuclear gluon distribution with heavy quarks}

\def\BA{\begin{eqnarray}}
\def\BE{\begin{equation}}
\def\BF{\begin{figure}[htb]}
\def\BT{\begin{table}[t]}
\def\EA{\end{eqnarray}}
\def\EE{\end{equation}}
\def\EF{\end{figure}}
\def\ET{\end{table}}

\newcommand{\beq}{\begin{equation}}
\newcommand{\eeq}{\end{equation}}
\newcommand{\beqn}{\begin{eqnarray}}
\newcommand{\eeqn}{\end{eqnarray}}

\newcommand\eps\varepsilon
\newcommand\euler{{\rm e}}
\newcommand\imag{{\rm i}}

\newcommand\unint{{\cal F}}
\newcommand\as{{\alpha_{\rm s}}}  
\newcommand\aem{{\alpha_{\rm em}}}

\def\sig{\sigma_{q\bar q}}

\def\fm{\,\mbox{fm}}
\def\GeV{\,\mbox{GeV}}

\def\lsim{\mathrel{\rlap{\lower4pt\hbox{\hskip1pt$\sim$}}
    \raise1pt\hbox{$<$}}}         
\def\gsim{\mathrel{\rlap{\lower4pt\hbox{\hskip1pt$\sim$}}
    \raise1pt\hbox{$>$}}}         

\begin{document}
\title{Probing the nuclear gluon distribution\\ with heavy quarks\thanks{Presented at ISMD04, Sonoma State U., Rhonert Park, CA, July~26 -- August~1, 2004}%
}
\author{J.~Raufeisen
\address{Institut f\"ur Theoretische Physik der Universit\"at, Philosophenweg 19,\\ 69120 Heidelberg, Germany,
{\tt J.Raufeisen@tphys.uni-heidelberg.de}}
}
\maketitle
\begin{abstract}
The color dipole formulation provides an intuitive picture of hard processes in high energy scattering.
Most importantly, this approach allows one to calculate nuclear effects %in such processes 
in a parameter-free way. I review the relation between the dipole approach and transverse momentum 
factorization and present numerical results for open heavy flavor production in proton-proton and proton-nucleus collisions. The cross section for heavy quark production off nuclei is known to reflect gluon shadowing, but
is also affected by higher twist effects and by finite coherence length effects. 
\end{abstract}
  
\section{Introduction}

Hadroproduction of heavy quarks at RHIC and LHC is of particular interest because this process allows one
to measure the poorly known gluon distribution of protons and especially of nuclei \cite{JC,Kirill}. So far, almost all 
our knowledge of the proton gluon density at low Bjorken-$x$ ($x_{Bj}$) is derived from deep inelastic scattering (DIS)
experiments at the Desy $ep$ collider HERA. However, since DIS measures the quark distribution, information about gluons
can be obtained only indirectly through the (linear) QCD evolution equations. 
These equations predict a strong rise of the gluon density at low $x_{Bj}$, simply because 
gluons are vector particles and are therefore most likely radiated with small momentum fraction.

Naturally, the gluon density cannot increase indefinitely because at some critical density, gluons will feel
each others proximity and recombine \cite{GLR}. Since the probability for recombination is proportional 
to the square of the gluon density, this saturation 
effect leads to non-linear corrections to the QCD evolution equations.
Gluon saturation has been the subject of intense theoretical and experimental investigations.

Nuclei have an advantage over protons for the study of saturation, since 
they have more gluons to start with. Therefore, non-linear effects in the evolution
of the nuclear gluon distribution should set in at much lower energy than for protons.
Moreover, large nuclei have an intrinsic hard scale, the so-called 
saturation scale $Q_s^2$, which is proportional
to the nuclear thickness $T_A(b)$. This
gives rise to the hope that saturation is calculable in perturbative QCD.

Non-linear effects are expected when $\alpha_sT_A(b)xG(x)\sim Q^2$, where $G(x)$ is the gluon
density of the target and $Q^2$ is the hardness of the probe that measures $G(x)$. (In the case of heavy quark production, $Q^2\approx m_Q^2$, where $m_Q$ is the mass of the heavy quark.)
In the
rest frame of the nucleus, this condition simply means that the mean free
path of a probe of transverse size $1/Q^2$ is smaller than the length of the medium.
Because of this intuitive interpretation of saturation as multiple scattering, it is often advantageous 
work in the 
target rest frame.  

In what follows, I shall employ the color dipole approach to discuss open charm and bottom production 
at RHIC and LHC. The dipole approach is formulated in the target rest frame, but all observables are 
Lorentz-invariant, of course. Since the dipole approach is most easily understood and most widely used in
DIS, I shall discuss the latter process first, before turning to the main subject of heavy quark 
hadroproduction.

\section{Deep Inelastic Scattering}

\begin{figure}[bt]
  \centerline{
  \scalebox{1}{\includegraphics{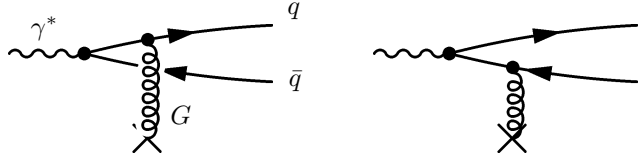}}}
\center{%\parbox[thb]{13cm}
{\caption{\em
  \label{fig:2graphs} Perturbative QCD graphs for DIS. 
  The virtual photon ($\gamma^*$) fluctuates into a (virtual)
$q\bar q$-pair
far before the target. The interaction with the target, which is denoted by 
gluon ($G$) exchange, can put this fluctuation on mass shell.
  }  
    }  }
\end{figure}

In the target rest frame, DIS
looks like pair creation in the gluon field of the target, 
see Fig.~\ref{fig:2graphs}. For simplification, I consider only the
case of a longitudinally polarized $\gamma^*$ 
with virtuality $Q^2$. The
transverse polarization does not contain any qualitatively new
physics. A straightforward calculation of the two Feynman diagrams
in Fig.~\ref{fig:2graphs} yields for the transverse momentum 
($\vec\kappa_\perp$)
distribution of the quark,
\beqn\nonumber\label{eq:kt}\lefteqn{
\frac{d\sigma\left(\gamma^*_Lp\to qX\right)}{d^2\kappa_\perp}=\frac{4\aem\as e_Q^2Q^2}{\pi}\int d\alpha\, \alpha^2\left(1-\alpha\right)^2}
\\ &\times&
\int\frac{d^2k_T}{k_T^2}\left(\frac{1}{\kappa_\perp^2+\eps^2}-
\frac{1}{(\vec\kappa_\perp-\vec k_T)^2+\eps^2}\right)^2
\unint(x,k_T^2),
\eeqn
where $\eps^2=\alpha\left(1-\alpha\right)Q^2+m_q^2$ and
$\alpha$ is the light-cone momentum fraction of the quark.  
The $q\bar q$-pair exchanges a
gluon with transverse momentum $\vec k_T$ with the target. The latter is
characterized by its unintegrated gluon density $\unint(x,k_T^2)$.
Note that Eq.~(\ref{eq:kt}) is also valid for heavy flavors.

%%%%%%%%%%%%%%%%%%%%%%%%%%%%%%%%%%%%%%%%%%%%%%%%%%%%%%%%%%%%%%%%%%%%%%%
%%%%%%%%%%%%%%%%%%%%%%%%%%%%%%%%%%%%%%%%%%%%%%%%%%%%%%%%%%%%%%%%%%%%%%%

In the dipole approach, a mixed representation is employed, that
treats the longitudinal ($\gamma^*$) direction in momentum space,
but the two transverse directions are described in coordinate 
({\em i.e.} impact parameter) space. After Fourier transforming Eq.~(\ref{eq:kt}),
one obtains
\beqn\label{eq:ptdis}\lefteqn{
\frac{d\sigma\left(\gamma^*_Lp\to qX\right)}{d^2\kappa_\perp}=
\frac{1}{2(2\pi)^2}\int d\alpha \int d^2{r}_1 d^2{r}_2 
\euler^{\imag\vec \kappa_\perp\cdot(\vec{r}_1-\vec{r}_2)}}\\
\nonumber&\times&
\Psi^L_{\gamma^*\to q\bar q}(\alpha,\vec{r}_1)
\Psi^{*L}_{\gamma^*\to q\bar q}(\alpha,\vec{r}_2)
%\\
%&\times&
\Bigl\{
\sigma_{q\bar q}({r}_1,x)
+\sigma_{q\bar q}({r}_2,x)
-\sigma_{q\bar q}(\vec{r}_1-\vec{r}_2,x)
\Bigr\}.
\eeqn
Since the $\vec r_i$ are conjugate variables to $\vec\kappa_\perp$,
one can interpret $\vec r_1$ as the transverse size of the $q\bar q$-pair
in the amplitude and $\vec r_2$ as the size of the pair in the complex
conjugate amplitude. 
%An expression similar to Eq.~(\ref{eq:ptdis})
%was also obtained in Ref.~\cite{Mueller99}.
These wavefunctions are simply the $\gamma^*\to
q\bar q$ vertex times the Feynman propagator for the quark line in
Fig.~\ref{fig:2graphs}, and can therefore be calculated in perturbation
theory. Explicit expressions can be found {\em e.g.} in \cite{green}.

The flavor independent dipole cross section $\sig$ in Eq.~(\ref{eq:ptdis})
carries all the information about the target.
It is related to the unintegrated gluon density by \cite{gBFKL}
\beq\label{eq:dipsec2}
\sig(r,x)=\frac{4\pi}{3}\int\frac{d^2k_T}{k_T^2}\as \unint(x,k_T)
\left\{1-\euler^{-\imag\vec k_T\cdot\vec r}\right\}.
\eeq
The color screening factor in the curly brackets in Eq.~(\ref{eq:dipsec2})
ensures that $\sig(r,x)$ vanishes $\propto r^2$ (modulo logs) at small
separations. %This seminal property of the dipole cross section is 
%known as {\em color transparency} \cite{ZKL,ct,bm}. 
The dipole cross
section cannot be calculated from first principles, but has to be determined
from experimental data. 

In the high energy limit, one can neglect the dependence of the gluon
momentum fraction $x$ on $\vec\kappa_\perp$ and integrate Eq.~(\ref{eq:ptdis})
over $\vec\kappa_\perp$ from $0$ to $\infty$. 
One then obtains a particularly simple formula
for the total cross section,
\beq\label{eq:totaldis}
\sigma_{\rm tot}\left(\gamma^*p\to X\right)=
\sum_{T,L}\int d\alpha \int d^2{r}\left|
\Psi^{T,L}_{\gamma^*\to q\bar q}(\alpha,\vec{r}_1)\right|^2
\sigma_{q\bar q}({r},x).
\eeq
This shows how the dipole formula Eq.~(\ref{eq:totaldis}) is related to 
transverse momentum factorization, Eq.~(\ref{eq:kt}).
With a phenomenological parametrization of $\sigma_{q\bar q}(r,x)$ (see Ref.~\cite{Bartels}), 
an
excellent description of HERA low $x_{Bj}$ data is achieved.

It was first realized in \cite{ZKL} that at high energies, color dipoles
with a well defined transverse separation are the eigenstates of the interaction, {\em i.e.}
the eigenvalues of the elastic amplitude operator
are given by the dipole cross section.
For that reason, it is particularly simple to calculate multiple scattering
effects in the dipole picture: at asymptotically high energy, 
the $q\bar q$ is formed long before the target, and its transverse size is frozen by Lorentz time
dilation. In that case, one can simply eikonalize $\sig(r)$, 
 \beq\label{eq:eikonal}
\widetilde\sigma^{A}_{q\bar q}(r) = 2\int d^2b
\left\{1-\exp\left[-{1\over2}\sig(r)T_A(b)\right]\right\}\ .
\label{bzk2}
 \eeq
At realistic energies, however, one has to take the finite lifetime of the $q\bar q$-pair into
account \cite{green}. Data on nuclear shadowing in DIS are then well reproduced \cite{krt}.

Note that even though a parametrization of $\sig(r,x)$ takes into account all higher Fock states
of the $\gamma^*$, eikonalization accounts only for the rescattering of the lowest Fock state, {\em i.e.}
the $q\bar q$-pair. Multiple scattering of higher Fock components containing gluons
is however especially important in the case of longitudinal photons and for heavy quarks.
These need to be treated separately, as will be explained in the following section.
 
\section{Hadroproduction of Heavy Flavors}

\begin{figure}[bt]
  \centerline{\scalebox{0.44}{\includegraphics{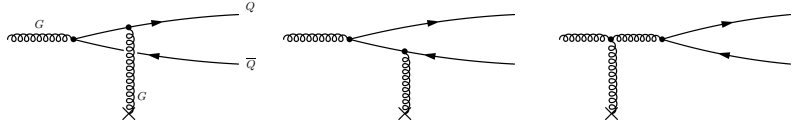}}}
    \center{%\parbox[b]{13cm}
    {\caption{\em
      \label{fig:3graphs}The three lowest order
graphs contributing to heavy quark
 production in the dipole approach.  
These graphs correspond to the gluon-gluon
fusion mechanism of heavy quark production in the parton model.}  
    }  }
\end{figure}

Eq.~(\ref{eq:totaldis}) is a special case of the general rule that
at high center of mass energies 
$\sqrt{s}$, 
the cross section for any reaction $a+N\to\{b,c,\dots\}X$ can be expressed
as convolution of the light-cone (LC) wavefunction for the transition $a\to\{b,c,\dots\}$
and the cross section for scattering the color neutral
$\{{\rm anti-}a,b,c\dots\}$-system off the target nucleon $N$. In the case of heavy quark production this means that the Feynman graphs in Fig.~\ref{fig:3graphs} can be written in the form \cite{npz,kt},
\beq\label{eq:all}
\sigma(GN\to \{Q\overline Q\} X)
=\int_0^1 d\alpha \int d^2{r} 
\left|\Psi_{G\to Q\overline Q}(\alpha,{r})\right|^2
\sigma_{q\bar q G}(\alpha,{r}),
\eeq 
where $\sigma_{q\bar qG}$ is
the cross section for scattering a color neutral quark-antiquark-gluon
system off a nucleon \cite{kt},
\beq\label{eq:qqG}
\sigma_{q\bar qG}(\alpha,{r})
=\frac{9}{8}\left[\sigma_{q\bar q}(\alpha{r})
+\sigma_{q\bar q}((1-\alpha){r})\right]
-\frac{1}{8}\sigma_{q\bar q}({r}).
\eeq
The dipole cross section $\sigma_{q\bar q}$ is the same as in DIS.
Again, the LC wavefunctions $\Psi_{G\to Q\overline Q}$ can be calculated perturbatively \cite{kt}.

%%%%%%%%%%%%%%%%%%%%%%%%%%%%%%%%%%%%%%%%%%%%%%%%%%%%%%%%%%%%%%%%%%%%%%%%%%%%%%
%%%%%%%%%%%%%%%%%%%%%%%%%%%%%%%%%%%%%%%%%%%%%%%%%%%%%%%%%%%%%%%%%%%%%%%%%%%%%

In momentum space, Eq.~(\ref{eq:all}) can be written in $k_\perp$-factorized form \cite{kt,catani}. Hence the dipole formulation is closely related to the approach of Ref.~\cite{Kirill}. However, the dipole approach takes into account only the finite transverse momentum of the target gluon. The relation of the dipole approach to the conventional parton model is explained in Ref.~\cite{JC}.

%%%%%%%%%%%%%%%%%%%%%%%%%%%%%%%%%%%%%%%%%%%%%%%%%%%%%%%%%%%%%%%%%%%%%%%%%%%%%%%%%%%
\begin{figure}[b]
  \centerline{\scalebox{0.39}{\includegraphics{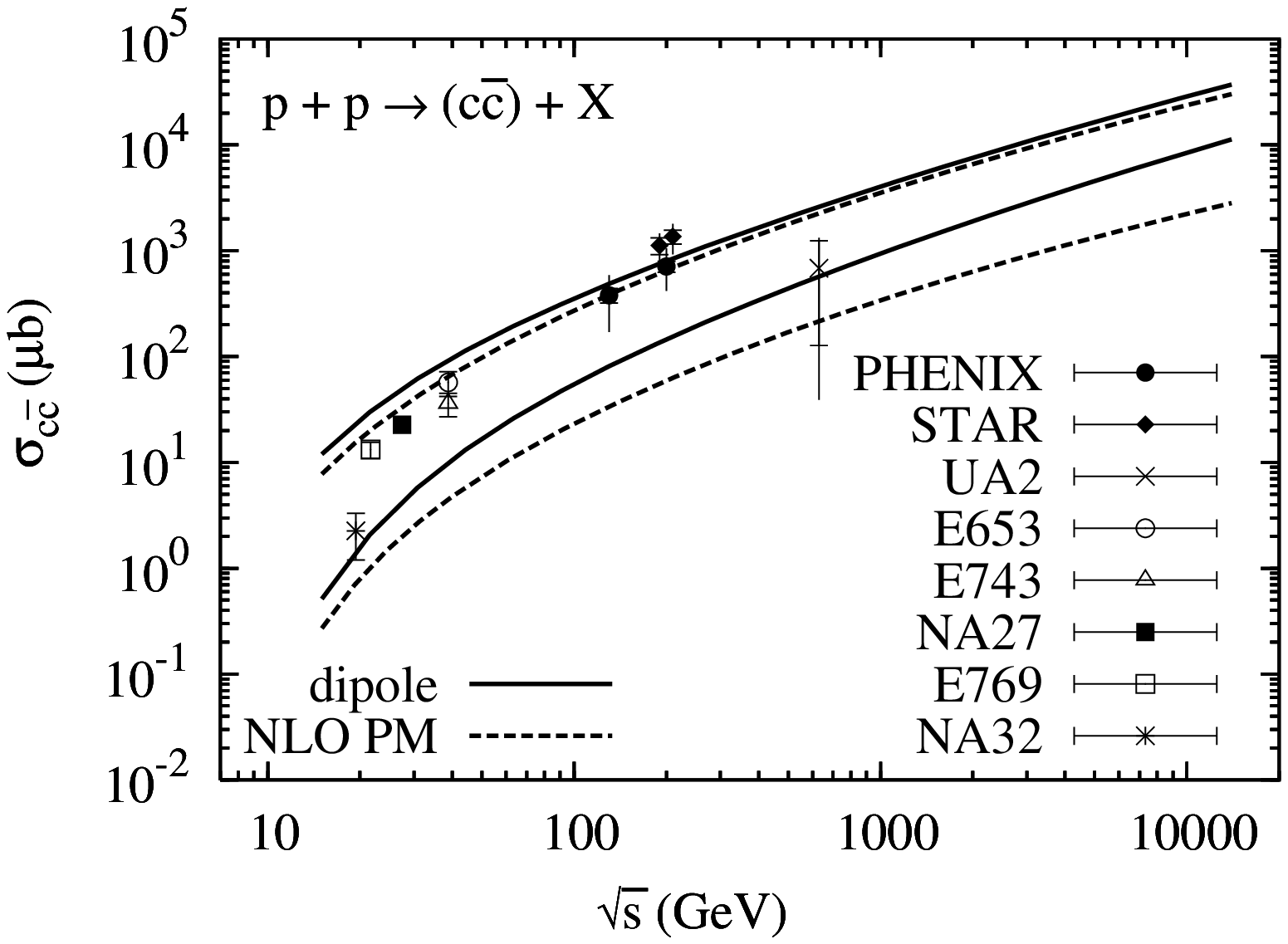}}
	      \scalebox{0.37}{\includegraphics{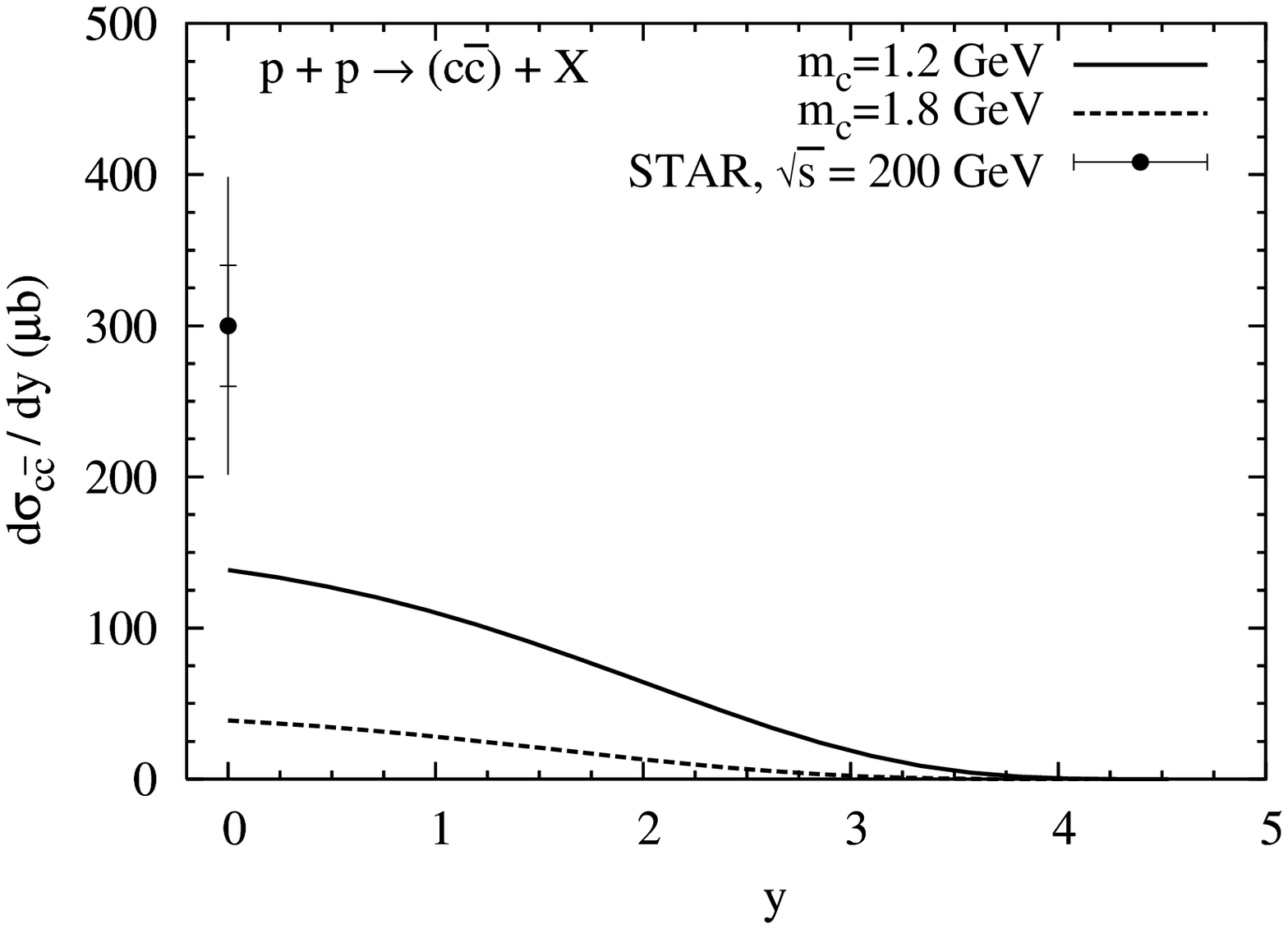}}}
\center{%\parbox[thb]{13cm}
{\caption{\em
      \label{fig:uncert}
Theory vs.\ data for open charm production. Left panel: vary\-ing free para\-meters in the dipole approach
and in the NLO
parton model (NLO PM) \cite{nlo}
gives rise to the uncertainties. 
Right panel: Single inclusive rapidity distribution in the dipole approach
for different values of the charm quark mass.
 }  
    }  }
\end{figure}
%%%%%%%%%%%%%%%%%%%%%%%%%%%%%%%%%%%%%%%%%%%%%%%%%%%%%%%%%%%%%%%%%%%%%%%%%%%%%%%%%%%%%

In order to calculate the cross section for heavy quark pair production
in proton-proton ($pp$) collisions,
Eq.~(\ref{eq:all}) has to be weighted with the projectile gluon density. Numerical results are shown in Fig.~\ref{fig:uncert}. Varying free parameters, such as the heavy quark mass $m_Q$, leads to the uncertainty represented by the space in between the curves. See Ref.~\cite{JC} for a complete discussion.
All data except the recent STAR measurement \cite{STAR} lie inside the uncertainty band. 
Note that one could increase the upper bound in Fig.~\ref{fig:uncert} (right) by allowing renormalization and factorization scales to be different.

Multiple scattering effects in nuclei become important at low $x_2$, when the lifetime of the $Q\overline Q$-pair becomes large enough to allow for more than one interaction with the target. There are two different sources of nuclear suppression for heavy quark production \cite{kt}. I discuss these now.

First, the $Q\overline Q$ pair in Fig.~\ref{fig:3graphs} can rescatter several times inside the target. Since the amplitude for interaction with a single nucleon is known and because dipoles are interaction eigenstates, all these rescatterings can be resummed by eikonalizing $\sigma_{q\bar q}$ as in Eq.~(\ref{eq:eikonal}). Note that since the typical size of the pair is ${r}^2\sim 1/m_Q^2$, double scattering behaves parametrically like $\Delta\sigma/\sigma\propto Q_s^2(x_2)/m_Q^2$. Though formally higher twist, nuclear suppression due to heavy quark rescattering is enhanced by a factor of the saturation scale in the target nucleus, $Q_s^2(x_2)\propto A^{1/3}$ and cannot be neglected for charm quarks. 

In addition, there is the leading twist gluon shadowing \cite{kst2}. One of the quarks (or the gluons) in Fig.~\ref{fig:3graphs} can radiate another gluon, which then propagates together with the $Q\overline Q$-pair through the nucleus. The magnitude of the rescattering correction is determined by the transverse distance ${r}_G$ this gluon can propagate from its parent quark. Unlike the size of the heavy quark pair, ${r}_G$ is limited only by nonperturbative QCD effects. In Ref.~\cite{kst2}, these nonperturbative effects were modeled for the case of DIS by an attractive interaction between the gluon and the quark it is radiated off. In order to explain the smallness of the Pomeron-proton cross section, a value ${r}_G\lsim 0.3\fm$ is needed, resulting in a rather weak gluon shadowing in DIS. I argued in
Ref.~\cite{QM} that the gluon propagation radius of $~0.3$~fm is determined by properties of the QCD vacuum and 
therefore process independent.

Gluon shadowing $R_G(x_2,b)$ at impact parameter $b$ is included in the calculation through the modified eikonal approximation,
\beq\label{eq:eik}
\sigma_{q\bar q}^A(x_2,{r})=2\int d^2b\left\{1-\exp\left(-\frac{1}{2}\sigma_{q\bar q}(x_2,{r})T_A(b)R_G(x_2,b)\right)\right\}.
\eeq
Note that the leading twist gluon shadowing has the effect of making the target nucleus more dilute, thereby reducing the saturation scale $Q_s^2$ and diminishing non-linear effects.

\begin{figure}[t]
\centerline{\scalebox{0.36}{\includegraphics{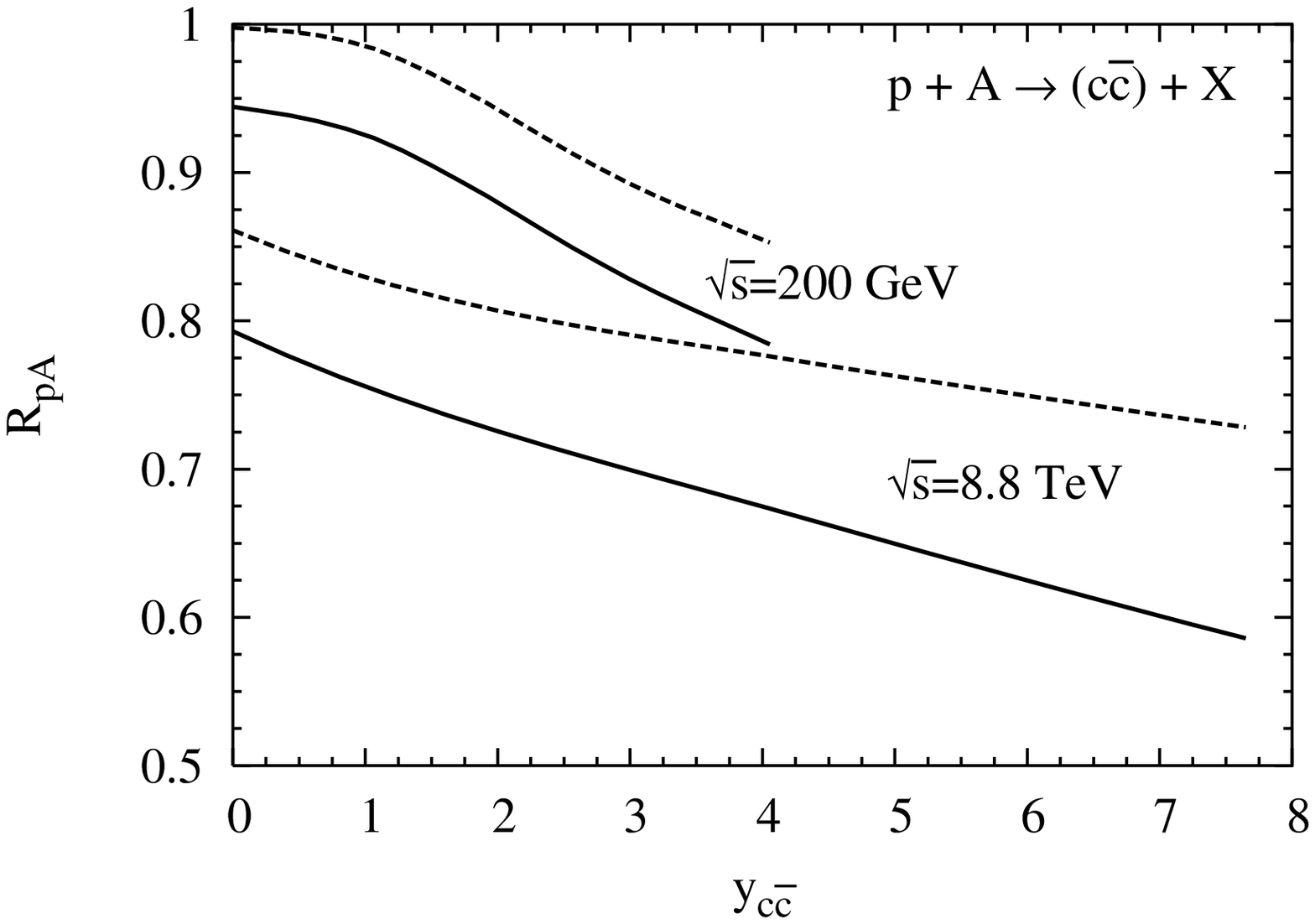}}
	      \scalebox{0.36}{\includegraphics{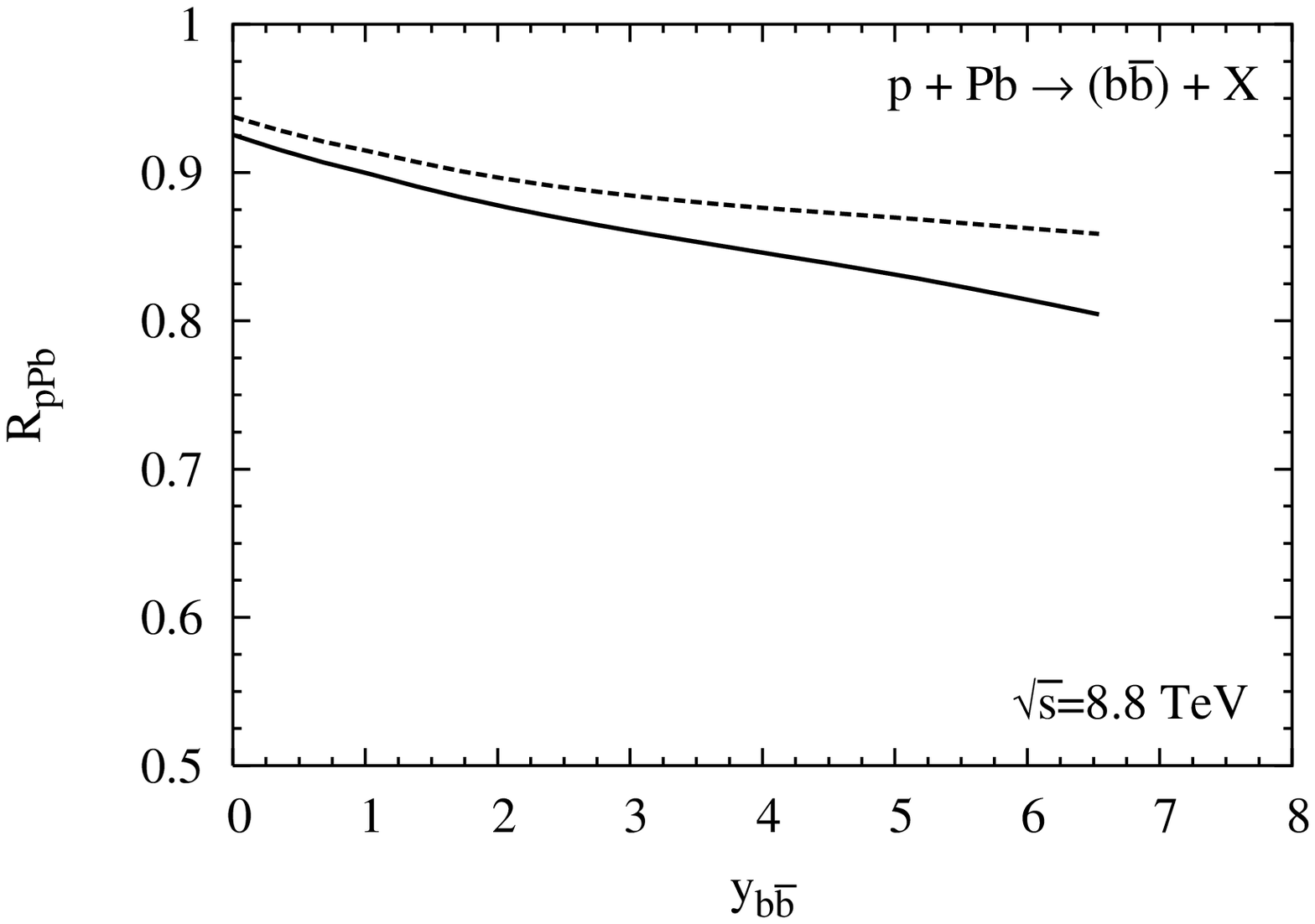}}}
  %\centerline{\scalebox{0.45}{\includegraphics{charm.eps}}}
    \center{%\parbox[b]{13cm}
    {\caption{\em
      \label{fig:charm} Rapidity dependence of nuclear suppression for open heavy flavor production in $pAu$ ($\sqrt{s}=200 \GeV$) and in $pPb$ ($\sqrt{s}=8.8$ TeV) collisions. Dashed curves show gluon shadowing only, while solid curves also include rescattering of the heavy quark pair.}  
    }  }
\end{figure}

Results are shown in Fig.~\ref{fig:charm}. The eikonal approximation assumes that the lifetime of the heavy quark pair (the coherence length $l_c\propto1/x_2$) is much larger than the nuclear radius $R_A$. This is certainly the case at the LHC ($\sqrt{s}=8.8$ TeV). For $y_{c\bar c}=0$ at RHIC, however, $l_c\sim R_A$ and a more sophisticated calculation using the Green's function technique developed in Ref.~\cite{green} would yield a slightly smaller suppression. 
Gluon shadowing sets in at smaller $x_2$ than quark shadowing, since the $Q\overline QG$-fluctuation has a much shorter lifetime than the $Q\overline Q$-pair alone because of the larger invariant mass of the $Q\overline QG$-state \cite{lc}. This is taken into account in this calculation. In particular there is no gluon shadowing at $y_{c\bar c}=0$ at RHIC. The rescattering of the heavy quark pair is a sizable contribution to the nuclear suppression of open charm production, especially at $\sqrt{s}=200\GeV$, where gluon shadowing is weak. 
%This is one reason why the observed suppression in $J/\Psi$ yields in $dAu$ collisions at RHIC cannot be explained by a modification of the nuclear gluon distribution alone. 
The rescattering of $b\bar b$-pairs at the LHC is expected to be only a small effect.

\section{Summary}

At high energies, hadroproduction of heavy quarks can be formulated in terms of the same
color dipole cross section as low $x_{Bj}$ DIS. 
Even though the dipole formulation is most intuitive for DIS, the cross section for
several processes can be 
expressed in terms of $\sig(r)$, {\em e.g.} Drell-Yan dilepton production \cite{krtj} and deeply 
virtual Compton scattering \cite{magno}.
It is not necessary that a $q\bar q$-dipole be present diagrammatically.
The dipole approach is closely related to transverse momentum factorization, the dipole cross section
being the Fourier transform of the unintegrated gluon density.

Using a successful phenomenological parameterization of $\sig(r,x)$ \cite{Bartels}, 
data on open charm hadroproduction in $pp$ are well described \cite{JC}, 
except for a recent STAR measurement \cite{STAR}.

The advantage of the dipole formulation is that it allows one to calculate nuclear effects without introducing
additional free parameters. All parameters are fitted to scattering off protons. 
The dipole approach takes into account both, rescattering of the $Q\overline Q$-pair as well as leading
twist nuclear gluon shadowing.
For open charm production, both of these effects are important, but for open bottom production
gluon shadowing dominates. An unambiguous extraction of nuclear gluon shadowing is therefore 
possible from a measurement of open bottom production at the LHC.

%\medskip
{\bf Acknowledgments: }{I thank the organizers of ISMD04 for inviting me to this stimulating meeting. I am indebted to Stan Brodsky, Boris Kopeliovich and Jen-Chieh Peng for valuable discussion. This research was supported in part under the Feodor Lynen Program of the Alexander von Humboldt Foundation and by BMBF under contract
06 HD 158.
}

\end{document}